\documentclass[twocolumn,aps,english,prl,nofootinbib]{revtex4}
\usepackage{mathrsfs}
\usepackage{graphicx}
\usepackage{hyperref}
\usepackage{amsmath, amssymb}
\usepackage{babel}
\usepackage{color}
\usepackage{slashed}

\newcommand{\boldtau}{\mbox{\boldmath $\tau$}}
\newcommand{\boldpi}{\mbox{\boldmath $\pi$}}

\begin{document}

\title{Relating hadronic CP-violation to higher-twist distributions}

\author{Chien-Yeah Seng$^{a,b}$}

\affiliation{$^{a}$INPAC, Shanghai Key Laboratory for Particle Physics and Cosmology, \\
	MOE Key Laboratory
	for Particle Physics, Astrophysics and Cosmology,  \\
	School of Physics and Astronomy, Shanghai Jiao-Tong University, Shanghai 200240, China\\
$^{b}$Helmholtz-Institut f\"ur Strahlen- und Kernphysik and Bethe Center for Theoretical Physics,\\
 Universit\"at Bonn, 53115 Bonn, Germany 	}

\date{\today}

\begin{abstract}

The nucleon sigma term of the isoscalar and isovector quark chromo-magnetic dipole moments are essential inputs for the determination of the CP-odd pion-nucleon couplings induced by quark chromo-electric dipole moments. We demonstrate that the former can be mapped to the third moment of the nucleon twist-three chiral-odd distribution functions $e^q(x)$ which are in principle measurable in semi-inclusive deep inelastic scattering processes. We perform a survey on existing model calculations as well as experimental data on $e^u(x)+e^d(x)$ and derive a predicted range for the isoscalar chromo-magnetic dipole moment sigma term.

\end{abstract}

\maketitle

The existence of a significant amount of CP-violation---much stronger than that supplied by the complex phase in the Cabibbo-Kobayashi-Maskawa (CKM) matrix---in fundamental interactions constitutes one of the three conditions \cite{Sakharov:1967dj} necessary to account for the observed baryon-antibaryon asymmetry in the universe \cite{Aghanim:2018eyx}. This triggers world-wide experimental searches of new CP-violating sources. They include, in the lepton sector, the Dirac phase in the Pontecorvo-Maki-Nakagawa-Sakata (PMNS) matrix that enters neutrino oscillations as well as the Majorana phases \cite{Pontecorvo:1967fh,Bilenky:1980cx} that can be probed in neutrinoless double beta decays. In the hadronic sector, special attention is paid to the permanent electric dipole moments (EDMs) of hadrons, nuclei and atoms.

The precise calculations of low-energy matrix elements of strongly-interacting bound states associated to the CP-violating observables are often crucial in translating the results of the above-mentioned searches into constraints on the CP-odd sector in the Beyond Standard Model (BSM) physics. These calculations are, however, very challenging due to the non-perturbative nature of Quantum Chromodynamics (QCD) at the hadronic scale. Although in some cases lattice QCD offers a convincing first-principle approach to the problem, the technical complexities of such calculations may greatly vary depending on the details of the desired matrix elements. An alternative approach is to relate the hadronic matrix elements of interest to experimental observables. A classic example of this kind is the relation between the nucleon tensor charges $\delta q$ and the first moment of the quark's transversity distribution function $h_1^q(x)$ \cite{Jaffe:1991kp,Jaffe:1991ra}.
The latter probes the difference in probability for a quark to be polarized parallel or anti-parallel to a transversely-polarized hadron and can be measured in experiment. The observation of such a relation is extremely important as it turns the purely theoretical challenge of the $\delta q$ calculation into an experimental problem. In fact, it stimulated a number of experiments to extract the tensor charges through semi-inclusive deep inelastic scattering (SIDIS), semi-inclusive $e^+e^-$ annihilation, and $\pi^0/\eta$-exclusive electroproduction 
\cite{Kang:2015msa,Radici:2015mwa,Goldstein:2014aja,Anselmino:2013vqa}. 

In this Letter we extend the idea above to the study of long-range CP-odd nuclear interactions induced by the dimension-5 quark chromo-electric dipole moment (cEDM) operators. These interactions are interesting because in many cases they are the main contributors to EDMs of multi-nucleon systems \cite{Engel:2013lsa,Yamanaka:2016umw,Yamanaka:2017mef,Chupp:2017rkp}. Chiral symmetry relates the CP-odd pion-nucleon couplings to nucleon matrix elements of the form  $\left\langle N\right|\bar{q}q\left|N\right\rangle$ and $\left\langle N\right|\bar{q}\sigma\cdot G q\left|N\right\rangle$ as we shall discuss later. The first matrix element is just the ordinary QCD sigma term which is studied extensively using lattice QCD and dispersion relation. The second, on the other hand, represents the nucleon matrix element of the quark chromo-magnetic dipole moment (cMDM) which is much harder, and its full lattice calculation is not yet available. 

We shall offer here a novel starting point to the problem by relating $\left\langle N\right|\bar{q}\sigma\cdot G q\left|N\right\rangle$ to a special class of experimental observables, namely, the chiral-odd twist-three distribution function $e^q(x)$ which can be probed in SIDIS experiments. This is interesting because it represents yet another nice interplay between three distinct branches of physics: (1) the precision frontier in BSM searches, (2) experimental studies of hadronic and nuclear structures and (3) lattice QCD. It points out another direction in the research of hadronic CP-violation apart from the conventional lattice or low-energy QCD model approach and provides extra motivation for the future improvement of experimental measurement of higher-twist observables. These experiments will also serve as a consistency check of the lattice result when the latter is available.

We start by considering a two-flavor QCD Lagrangian with the inclusion of a non-zero $\theta$-term as well as the quark cEDM operators:
\begin{equation}\label{eq:LQCD}
\mathcal L =\mathcal{L}_\mathrm{QCD}-\frac{ig_s}{2} \bar Q \sigma^{\mu\nu} G_{\mu\nu}
\tilde{d}_{CE}  \gamma_5  Q
-\frac{\bar{\theta}}{32\pi^2}G^a_{\mu\nu}\tilde{G}^{a\mu\nu},
\end{equation}
with $Q=(u,d)^T$ the isospin doublet quark field, $G^{a}_{\mu\nu}$ the gluon field strength tensor, $\tilde{G}^a_{\mu\nu}\equiv (1/2)\varepsilon_{\mu\nu\alpha\beta}G^{a\alpha\beta}$ its dual tensor, $\bar{\theta}$ the QCD-$\theta$ term and $\tilde{d}_{CE}=\mathrm{diag}(\tilde{d}_u,\tilde{d}_d)$ the cEDM coupling matrix. Sources of explicit chiral symmetry breaking (CSB) in the Lagrangian are the quark mass matrix $\mathcal{M}=\mathrm{diag}(m_u,m_d)$ and the cEDM matrix $\tilde{d}_{CE}$. Upon performing an anomalous $\mathrm{U(1)}_A$-rotation followed by a non-anomalous $\mathrm{SU(2)}_A$-rotation that aligns the vacuum \cite{deVries:2012ab,Bsaisou:2014oka,Dashen:1970et,Baluni:1978rf,Bhattacharya:2015rsa,Seng:2016pfd}, one absorbs all the CP-odd interactions into two explicit CSB terms in the Lagrangian:
\begin{equation}\label{eq:LQCDrotate}
\mathcal L_\mathrm{CSB} =-\bar{Q}_R\mathfrak{M}Q_L+\frac{ig_s}{2} \bar Q_R \sigma^{\mu\nu} G_{\mu\nu}\tilde{d}_{CE}Q_L+\mathrm{H.C}.
\end{equation}
among which the complex quark mass matrix $\mathfrak{M}$ is defined as $\mathfrak{M}\equiv\mathcal{M}+im_*(\bar{\theta}-\bar{\theta}_\mathrm{ind})+r\tilde{d}_{CE}$
where 
\begin{eqnarray}
m_*&=&\frac{\bar{m}(1-\varepsilon^2)}{2},\:\:\:\:
r\:=\:\frac{1}{2}\frac{\langle 0 | \bar{Q}g_s\sigma_{\mu\nu}G^{\mu\nu} Q | 0 \rangle}{\langle 0 | \bar{Q} Q | 0 \rangle},\nonumber\\
\bar{\theta}_\mathrm{ind}&=&r\mathrm{Tr}\left[\mathcal{M}^{-1}\tilde{d}_{CE}\right].
\end{eqnarray}
Here we have introduced the average quark mass $\bar{m}=(m_u+m_d)/2$ and the relative quark mass difference $\varepsilon=(m_d-m_u)/(2\bar{m})$. Similarly we shall also define the isoscalar and isovector cEDM constants $\tilde{d}_{0,3}\equiv(\tilde{d}_u\pm \tilde{d}_d)/2$. Notice that if the Peccei-Quinn (PQ) mechanism \cite{Peccei:1977hh} is at work, then $\bar{\theta}$ relaxes to $\bar{\theta}_{\mathrm{ind}}$ and simplifies the expression of $\mathfrak{M}$ above.

In a low-energy effective theory of hadrons, the presence of the above-mentioned CP-violating sources induces long-range CP-violating nuclear forces through the following pion-nucleon couplings:
\begin{equation}
\mathcal L_N =- \frac{\bar g_0}{2 F_{\pi}} \bar N \boldtau\cdot \boldpi N  - \frac{\bar g_1}{2 F_{\pi}} \pi_0 \bar N  N  - \frac{\bar g_2}{2 F_{\pi}} \pi_0 \bar N  \tau^3 N + \ldots,
\end{equation}
where $N=(p,n)^T$ is the nucleon isospin doublet and $F_\pi\approx 92.2$ MeV is the pion decay constant. The construction of numerically-precise relations between $\bar{g}_I$ and $\tilde{d}_i$ is currently a central question to both the precision frontier and the hadron physics community. It is well-known that these couplings could be related to the nucleon mass shifts induced by the quark masses and the cMDM operators \cite{Pospelov:2001ys,Bsaisou:2014oka,deVries:2015gea,Bsaisou:2012rg,Mereghetti:2010tp,deVries:2015una,Seng:2016pfd} through chiral symmetry. In particular, Ref. \cite{deVries:2016jox} suggests the following form of matching (PQ symmetry is assumed for simplicity):
\begin{equation}
\bar g_0=\tilde d_0\left(\sigma^3_C+\frac{r\sigma^3}{\bar{m}\varepsilon}\right),\:\:
\bar g_1 =-2 \tilde d_3 \left(\sigma^0_C-\frac{r\sigma^0}{\bar{m}}\right),\label{eq:matching}
\end{equation}
($\bar{g}_2$ is always of higher order and is neglected) where we define the isoscalar and isovector QCD sigma terms $\sigma^{0,3}$ and cMDM sigma terms $\sigma^{0,3}_C$ of a proton state $\left|P\right\rangle$ (with momentum $P^\mu$) as follows\footnote{We choose the normalization of the state as $\left\langle P|P'\right\rangle=(2\pi)^32E_P\delta^3(\vec{P}-\vec{P}')$. Here we also point out two typos in Eq. (34) of Ref. \cite{deVries:2016jox}: there should be a factor of 2 and -2 multiplied to $\langle p|\bar{q}\tau_3q|p\rangle$ and $\langle p|g_s\bar{q}\sigma_{\mu\nu}G^{\mu\nu}\tau_3 q|p\rangle$ respectively.}:
\begin{eqnarray}\label{eq:fh_matrix_elements}
\sigma^0&\equiv&\frac{\bar{m}}{2m_N}\left\langle P\right|\bar{Q}Q\left|P\right\rangle,\:\:\:
\sigma^3\equiv\frac{\bar{m}\varepsilon}{m_N}\left\langle P\right|\bar{Q}\tau_3Q\left|P\right\rangle \nonumber\\
\sigma^0_C&\equiv&\frac{1}{4m_N}\left\langle P\right|\bar{Q}g_s\sigma^{\mu\nu}G_{\mu\nu}Q\left|P\right\rangle \nonumber\\
\sigma^3_C&\equiv&-\frac{1}{2m_N}\left\langle P\right|\bar{Q}g_s\sigma^{\mu\nu}G_{\mu\nu}\tau_3Q\left|P\right\rangle 
\end{eqnarray} 
with $m_N$ the nucleon mass.
The relations in Eq. \eqref{eq:matching} are preserved exactly by one-loop corrections in chiral perturbation theory. The matchings are violated by $O(p^4)$ counterterms but the amount of violation is in general not larger than 10\%. 

It is instructive to write Eq. \eqref{eq:matching} as
\begin{equation} \bar{g}_I=\bar{g}_I\large|_\mathrm{dir}+\bar{g}_I\large|_\mathrm{vac},
\end{equation} 
i.e. to split $\bar{g}_I$ into the sum of ``direct" and ``vacuum alignment" contribution, which correspond to the first and second term at the right side of Eq. \eqref{eq:matching} respectively; the direct contribution depends on $\sigma_C^{0,3}$ while the vacuum alignment contribution depends on $\sigma^{0,3}$ as well as the vacuum condensate ratio $r$ and the current quark masses. Parameters in $\bar{g}_I\large|_{\mathrm{vac}}$ are rather extensively studied: for instance, simulations with $N_f=2$ provide numerical estimations for the isoscalar and isovector quark mass parameters and sigma terms: $\bar{m}\approx 3.6$ MeV, $\varepsilon\approx 0.33$, $\sigma^0\approx37$ MeV and $\sigma^3\approx 2.9$ MeV \cite{Bali:2012qs,deDivitiis:2013xla,Aoki:2016frl}\footnote{The quark mass parameters are evaluated at $\mu=2$ GeV whereas the sigma terms are scale-invariant.} which are also confirmed by more recent lattice calculations \cite{Bali:2016lvx,Alexandrou:2017qyt,Yamanaka:2018uud}. The isoscalar sigma term $\sigma^0$ may also be extracted from low-energy $\pi N$-scattering, but the outcomes are in general much larger than the lattice result (see \cite{RuizdeElvira:2017stg} and references therein). Meanwhile a rough estimate of the vacuum condensate ratio $r$ may be obtained from the QCD sum rule: $r\approx0.4$ GeV$^2$ \cite{Belyaev:1982cd,Pospelov:2001ys}. It however drops out when we construct the ratio between the vacuum alignment contribution to $\bar{g}_0$ and $\bar{g}_1$:
\begin{equation}
\left.\frac{\bar{g}_0}{\bar{g}_1}\right|_{\mathrm{vac}}=\frac{\sigma^3}{2\varepsilon\sigma^0}\frac{\tilde{d}_0}{\tilde{d}_3}\approx 0.12\frac{\tilde{d}_0}{\tilde{d}_3}.\label{eq:g0g1ratio}
\end{equation}
That is, if we neglect the direct contribution and assume that $\tilde{d}_0\sim\tilde{d}_3$, then one may conclude that $\tilde{g}_1\gg \tilde{g}_0$.

With that said, we still cannot claim to have a complete understanding of $\bar{g}_I$ without knowing the precise values of the cMDM sigma terms $\sigma^{0,3}_C$ that show up in $\bar{g}_I\large|_\mathrm{dir}$. Unlike the QCD sigma terms,  the lattice study of cMDM sigma terms is still in progress and the final result is not yet available \cite{LATT2018}. As mentioned earlier, we shall derive here a matching between $\sigma^{0,3}_C$ and the chiral-odd twist-three distribution function $e^q(x)$. We start by reviewing the basic properties of $e^q(x)$ of which details can be found in \cite{Efremov:2002qh} and references therein. First, for a proton state $|P\rangle$, $e^q(x)$ can be written as the matrix element of a quark bilinear with light-cone separation \cite{Jaffe:1991kp,Jaffe:1991ra}:
\begin{equation}
e^q(x)=\frac{1}{2m_N}\int\frac{d\lambda}{2\pi}e^{i\lambda x} \langle P | \bar{q}(0)[0,\lambda n]q(\lambda n) | P \rangle .
\end{equation}
where $n^\mu$ is a basis vector on the light cone and $[0,\lambda n]$ is the gauge link operator. This function is non zero at $-1\leq x\leq 1$ and satisfies $e^q(-x)=e^{\bar{q}}(x)$. The scalar bilinear operator $\bar{q}(0)[0,\lambda n]q(\lambda n)$ could be decomposed by the mean of operator identity \cite{Balitsky:1987bk,Braun:1989iv,Belitsky:1997zw,Kodaira:1998jn} which leads to the decomposition of $e^q(x)$ in terms of the
 ``singular", ``pure twist-three" and ``quark mass" terms:
 \begin{equation}
 e^q(x)=e^q_{\mathrm{sing}}(x)+e^q_{\mathrm{tw3}}(x)+e^q_\mathrm{mass}(x)
 \end{equation}
where the explicit form of each term can be found in Ref. \cite{Efremov:2002qh}.
For later convenience, we shall also define the (n+1)-th Mellin moment of a distribution function $F(x)$ as
\begin{equation}
F_n\equiv\int^1_{-1}dx x^n F(x).
\end{equation}
Here the integral over $x$ ranges from -1 to 1 so it includes simultaneously the effect of the parton and anti-parton.

We are particularly interested in the third moment of $e^q(x)$ that is contributed only by the quark mass term and the pure twist-three term:  \cite{Balitsky:1996uh,Belitsky:1997zw,Koike:1996bs,Burkardt:2008ps}: $e_2^q=e^q_{2,\mathrm{mass}}+e^q_{2,\mathrm{tw3}}$ where
\begin{eqnarray}
e^q_{2,\mathrm{tw3}}&=&\frac{1}{4m_N(P^{+})^2}\sum_{i=1}^2\langle P|\bar{q}(0)\sigma^{+i}g_sG^{+i}(0)q(0) | P \rangle\nonumber\\
e^q_{2,\mathrm{mass}}
&=&\frac{m_q}{m_N}f^q_1.
\end{eqnarray} 
Here the light-cone components of a four-vector $a^\mu$ are defined as $a^\pm=(1/\sqrt{2})(a^0\pm a^3)$. Notice that the third moment of $e_{\mathrm{mass}}^q$ is related to the second moment of the ordinary unpolarized quark distribution function $f^q(x)$. We shall now argue that the third moment of $e^q(x)$ is contributed mainly by $e^q_{\mathrm{tw3}}(x)$. First of all, $e^q_{2,\mathrm{mass}}$ receives a current quark mass suppression: $\bar{m}/m_N\sim 4\times 10^{-3}$; at the same time, both experiments and lattice simulations suggest $f^{u,d}_1\sim 10^{-1}$ at the hadron scale \cite{Deka:2008xr}. These together give $e^q_{2,\mathrm{mass}}\sim 10^{-3}-10^{-4}$. On the other hand, since $e^q_2$ is boost-invariant in the $z$-direction, we may work in the proton rest frame where $P^+=m_N/\sqrt{2}$. The order of magnitude of $e^q_{2,\mathrm{tw3}}$ can be roughly estimated using na\"{i}ve dimensional analysis (NDA) \cite{Weinberg:1989dx,Manohar:1983md}:
\begin{equation}
e^q_{2,\mathrm{tw3}}\sim\frac{1}{2m_N^3}\times\frac{\alpha_s}{4\pi}\Lambda_\chi^3\label{eq:NDA}
\end{equation}
where $\alpha_s=g_s^2/4\pi$ and $\Lambda_\chi\sim 1$ GeV is the so-called CSB scale. To get a feeling, we take $\alpha_s\approx 0.5$ at $\mu=1$ GeV in $\overline{\mathrm{MS}}$-scheme \cite{Deur:2016tte}; that gives $e^q_{2,\mathrm{tw3}}\sim 10^{-2}$ which is at least an order of magnitude larger than $e^q_{2,\mathrm{mass}}$. Therefore, it is reasonable to assume that $e^q_2$ is dominated by the pure twist-three contribution:
\begin{equation}
e^q_2\approx e^q_{2,\mathrm{tw3}}.\label{eq:e2e2tw3}
\end{equation}
Finally, the renormalization group (RG) evolution of $e^q(x)$ and its moments were studied in several papers \cite{Belitsky:1997zw,Balitsky:1996uh,Koike:1996bs}. 
For the third moment which is particularly important for us, we shall adopt an improved evolution formula including the $1/N_c^2$ corrections \cite{Koike:1996bs} in the chiral limit:
\begin{equation}\label{eq:e2evolve}
e^q_2(\mu)=\left(\frac{\alpha_s(\mu)}{\alpha(\mu_0)}\right)^{6.11/b}e^q_2(\mu_0).
\end{equation}
where $b=(11N_c-2N_f)/3$.

Now we shall demonstrate that the third moment of $e^q(x)$ is connected to the cMDM sigma terms defined in Eq. \eqref{eq:fh_matrix_elements}. This can be seen by considering the following parameterization of the (spin-averaged) $\bar{q}\sigma\cdot G q$ matrix element:
\begin{eqnarray}\label{eq:munumatrix}
\langle P|\bar{q}(0)g_sG^{\alpha\mu}(0)\sigma_\alpha^{\:\:\nu} q(0) | P \rangle&=&A^qm_N(m_N^2g^{\mu\nu}-P^\mu P^\nu)\nonumber\\
&&+B^qm_NP^\mu P^\nu,
\end{eqnarray}
where $A^q$ and $B^q$ are dimensionless, scale-dependent invariant matrix elements. The cMDM sigma term obviously depends on both $A^q$ and $B^q$:
\begin{equation}\label{eq:cMDMsigmaAB}
\langle P|\bar{q}(0)g_sG^{\alpha\mu}(0)\sigma_{\alpha\mu} q(0) | P \rangle=3A^qm_N^3+B^qm_N^3.
\end{equation}
On the other hand, $e^q_{2,\mathrm{tw3}}$ depends on another combination of $A^q$ and $B^q$:
\begin{equation}e^q_{2,\mathrm{tw3}}=\frac{A^q-B^q}{4}.
\end{equation}
Combining with the approximation in Eq. \eqref{eq:e2e2tw3} we thus obtain:
\begin{eqnarray}
\sigma^0_C&\approx&m_N^2\left(3(e_2^u+e_2^d)+B^u+B^d\right)\nonumber\\
\sigma^3_C&\approx&-2m_N^2\left(3(e_2^u-e_2^d)+B^u-B^d\right).\label{eq:cMDMmatcheq2}
\end{eqnarray}

This is the central result of the Letter: the cMDM sigma terms $\sigma^{0,3}_C$ are related to the third moment of $e^{u,d}(x)$, barring the two unknown constants $B^{u,d}$ that vanish in the non-relativistic (NR) limit. The latter can be seen by working in the nucleon's rest frame (i.e. $P^\mu=(m_N,\vec{0})$) and realizing that in the NR limit the quark bilinear $\bar{q}\sigma_\alpha^{\:\:\nu}q$ is non-zero only when $\alpha,\nu\neq 0$. That is, the right side of Eq. \eqref{eq:munumatrix} must be zero when $\mu=\nu=0$, and this can be achieved only if $B_q=0$. It is well-known that the symmetry relations obtained from NR quark models are identical to those implied by spin-flavor symmetry, which is a direct consequence of the large $N_c$-expansion \cite{Dashen:1994qi}. Therefore, the terms $B^q$ are subdominant and we shall neglect them in our numerical analysis henceforth. Under this assumption one could determine the cMDM sigma terms $\sigma^{0,3}_C$ through the precise experimental measurement of $e^{u,d}(x)$ in analogy to the acquirement of the nucleon tensor charge $\delta q$ from the transversity distribution function $h_1^q(x)$. 

Before diving into experiments, it is instructive to first gain some insights from various QCD models such as the bag model \cite{Jaffe:1991ra,Signal:1996ct}, spectator model \cite{Jakob:1997wg}, chiral quark-soliton model ($\chi$QSM) \cite{Schweitzer:2003uy,Ohnishi:2003mf,Cebulla:2007ej} and light-front constituent quark model (LFCQM)\cite{Lorce:2014hxa,Lorce:2016ugb,Pasquini:2018oyz}. The shape of $e^q(x)$ is plotted for each model which allows us to deduce the third moment of the isosinglet combination $e^u(x)+e^d(x)$ and evolve it to $\mu=1$ GeV, as summarized in Table \ref{tab:models}. Note that the outcomes are consistent with the NDA in Eq. \eqref{eq:NDA} which is reassuring. With these results, we deduce the implied isosinglet cMDM sigma term $\sigma^0_C$ using Eq. \eqref{eq:cMDMmatcheq2} (neglecting $B^{u,d}$). This provides us a range of model-predicted values of $\sigma^0_C$ as $(0.085-0.29)$ GeV$^2$.  We may use this to compare the relative importance between the ``direct" and ``vacuum alignment" contribution to $\bar{g}_1$:
\begin{equation}
\frac{\bar{g}_1|_{\mathrm{dir}}}{\bar{g}_1|_{\mathrm{vac}}}=-\frac{\bar{m}}{\sigma^0}\frac{\sigma^0_C}{r}\approx-0.63\left(e_2^u+e_2^d\right)=-(0.02-0.07),
\end{equation}
where we have taken the lattice value for $\bar{m}$, $\sigma^0$ and the sum-rule estimation of $r$ as described above Eq. \eqref{eq:g0g1ratio};
the outcome implies that the ``vacuum alignment" contribution to $\bar{g}_1$ dominates over the direct contribution. The readers should however be alerted that such conclusion relies critically on the values of $\bar{m}$, $\sigma^0$ and $r$; for instance, if for some reason the actual value of the vacuum ratio $r$ is a few times smaller than the sum-rule prediction, then $\bar{g}_1|_{\mathrm{dir}}$ could turn out to be comparable to $\bar{g}_1|_{\mathrm{vac}}$.

\begin{table}
	
	\begin{centering}
		\begin{tabular}{|c|c|c|}
			\hline 
			&$e_{2}^{u}+e_{2}^{d}$&$\sigma^0_C\cdot\mathrm{GeV}^{-2}$\tabularnewline
			\hline 
			\hline 
			Bag Model\cite{Jaffe:1991ra} & 0.032 &0.085 \tabularnewline
			\hline 
			Spectator Model\cite{Jakob:1997wg} & 0.042 &0.11\tabularnewline
			\hline 
			$\chi$QSM\cite{Cebulla:2007ej} & 0.063&0.17\tabularnewline
			\hline 
			LFCQM\cite{Pasquini:2018oyz,Pasquiniprivate} & 0.11&0.29\tabularnewline
			\hline 
		\end{tabular}
		\par\end{centering}
	\caption{\label{tab:models}Model predictions for the third moment of $e^{u}(x)+e^{d}(x)$ 
		in a proton evolved to $\mu=1$ GeV using Eq. \eqref{eq:e2evolve}, and the implied value for $\sigma^0_C$ at the same scale using Eq. \eqref{eq:cMDMmatcheq2} with the subdominant terms $B^{u,d}$ neglected.}
	
\end{table}

Finally, we may explore the implications on $\sigma_C^0$ from current experimental data.
We find two existing articles that attempted for the extraction of $e^q(x)$ from experiment: Refs. \cite{Efremov:2002ut,Courtoy:2014ixa} among which only the former is published in a peer-reviewed journal, and therefore we shall use it to obtain an estimate of $e_2^u+e_2^d$. In Ref. \cite{Efremov:2002ut}, the combination $e(x)\equiv e^u(x)+(1/4)e^{\bar{d}}(x)$ at the scale $Q^2=1.5$ GeV$^2$ was extracted from the azimuthal asymmetry $A_{LU}$ in the SIDIS process $ep\rightarrow e\pi^+X$ which was measured by the CLAS Collaboration \cite{Avakian:2003pk}. Such extraction required the knowledge of the Collins fragmentation function $H_1^\perp$ which was deduced from the HERMES data \cite{Efremov:2002td,Airapetian:1999tv,Airapetian:2001eg}.

It is non-trivial to translate the outcome in Ref. \cite{Efremov:2002ut} into our desired third moment of $e^u(x)+e^d(x)$ due to two reasons: (1) the measured combination $e(x)$ is not isoscalar and (2) there are altogether only four measured data points which lie within $0.18<x<0.37$. Therefore, extra assumptions are needed in order to extract the most information out of it. First, we assume $e^d(x)\approx e^u(x)$ that holds in the large $N_c$ limit \cite{Efremov:2002qh} (which implies $\sigma_C^3\ll\sigma_C^0$). Second, we adopt a simple Gaussian-like parameterization of $e^u(x)$:
\begin{equation}
e^u(x)=A\exp\left\{-\frac{(x-x_0)^2}{2\sigma^2}\right\}\label{eq:eupara}
\end{equation}
which qualitatively describes most of the model predictions. The parameters $\{A,x_0,\sigma\}$ are to be fitted to experimental data. 

\begin{figure}
	\begin{centering}
		\includegraphics[scale=0.4]{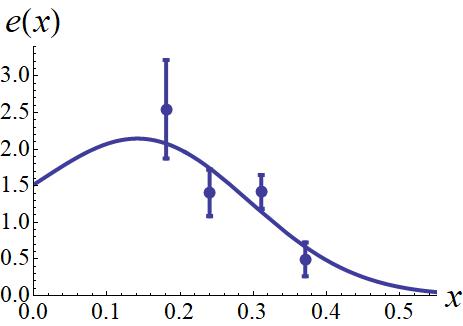} 
		\par\end{centering}
	\caption{\label{fig:fit}Simple fit to the CLAS result of $e(x)=e^u(x)+(1/4)e^{\bar{d}}(x)$ \cite{Efremov:2002ut} using $x_0=0.15$, $A=2.08$ and $\sigma=0.15$.}
\end{figure}

Our fitting proceeds as follows. First, we fit the four data points using the parameterization in Eq. \eqref{eq:eupara} which returns a best-fitted value of $x_0\approx 0.15$, in rough agreement with the phenomenological model predictions. Next we fix $x_0=0.15$ and repeat the fit to obtain the two remaining parameters: $A=2.08\pm 0.54$ and $\sigma =0.15\pm 0.03$; the fitted curve is shown in Fig. \ref{fig:fit}. With these we may compute the third moment of $e^u(x)+e^d(x)$ by varying $A$ and $\sigma$ within their respective allowed regions, which then gives $0.03<e^u_2+e^d_2<0.13$ at $Q^2=1.5\:\mathrm{GeV}^2$. With our master formula \eqref{eq:cMDMmatcheq2}, it implies:
\begin{equation}
\sigma_C^0=(0.08-0.34)\:\mathrm{GeV}^2,\:\:Q^2=1.5\:\mathrm{GeV}^2,
\end{equation}
consistent with the model-predicted range, which is not surprising because the fitting function \eqref{eq:eupara} itself is model-inspired.
The result above is obviously preliminary, but it points towards the possibility of a much more precise determination of the cMDM sigma terms with the future accumulation of more data points of $e^q(x)$, especially in the large-$x$ region that weights more in the calculation of its third moment. 

In summary, we establish a connection between the chiral-odd twist-three distribution functions $e^q(x)$ and the cMDM sigma terms $\sigma_C^{0,3}$ that are important in the discussion of long-range hadronic CP-violation and were once thought to be only obtainable by performing non-perturbative calculations such as lattice QCD. The relation is obtained by separating $e^q(x)$ into the ``singular", ``pure twist-three" and ``quark mass" terms and investigating the structure of their third Mellin moment. There are two major assumptions in arriving at a practically useful matching: (a) $e_2^q\approx e_{2,\mathrm{tw3}}^q$ in Eq. \eqref{eq:e2e2tw3} and (b) $B^q\rightarrow 0$ in Eq. \eqref{eq:cMDMmatcheq2}. While (a) is more justifiable, the assumption (b) may result in a systematic error of order $\mathcal{O}(1/N_c)$; one should however be reminded that a determination of the cMDM sigma terms with a (20-30)\% accuracy will already represent a non-trivial achievement at this stage. The function $e^q(x)$ can be probed, for instance, through the measurement of the azimuthal asymmetry $A_{LU}$ of a single-hadron SIDIS $\vec{l}N\rightarrow l'hX$ or a di-hadron SIDIS $\vec{l}N\rightarrow l'h_1h_2X$ between a longitudinally-polarized lepton and an unpolarized nucleon target. Future improvements in the precision of such measurements covering a wider region of $x$ will therefore bring benefits not only to the understanding of nuclear structure, but also to the precision frontier in BSM searches and to the lattice QCD community.

\bigskip

The author thanks Jordy de Vries, Andrea Signori, Shuai Zhao and Yong Zhao for many inspiring discussions. This work is supported in part by the National Natural Science Foundation of China (NSFC) under Grants Nos.11575110, 11655002, 11735010, Natural Science Foundation of Shanghai under Grants No.~15DZ2272100 and No.~15ZR1423100, by Shanghai Key Laboratory for Particle Physics and Cosmology, by Key Laboratory for Particle Physics, Astrophysics and Cosmology, China Ministry of Education, by the DFG (Grant No. TRR110)
and the NSFC (Grant No. 11621131001) through the funds provided to the Sino-German CRC 110 ``Symmetries and the Emergence of Structure in QCD". The author also appreciates the support through the Recruitment Program of Foreign Young Talents from the State Administration of Foreign Expert Affairs, China and by the Alexander von Humboldt Foundation through the Humboldt Research Fellowship.

\bibliography{gpiDIS_ref}

\end{document}